\documentclass[12pt,a4paper,oneside]{article}
\usepackage{graphicx,sectsty,amssymb,amsmath,dsfont,slashed}
\usepackage[linkcolor={blue},citecolor={red},colorlinks=true]{hyperref}
\usepackage[margin=2cm]{geometry}
\newcommand{\beq}{\begin{equation}}
\newcommand{\eeq}{\end{equation}}
\newcommand{\yut}{(Y_u Y_u^\dagger)_{\slashed{\mathrm{tr}}}}
\newcommand{\ydt}{(Y_d Y_d^\dagger)_{\slashed{\mathrm{tr}}}}
\newcommand{\au}{\mathcal{A}_u}
\newcommand{\ad}{\mathcal{A}_d}
\newcommand{\had}{\mathcal{\hat A}_d}
\newcommand{\hadp}{\mathcal{\hat A}'_d}
\newcommand{\hadn}{\mathcal{\hat A}_d^n}
\newcommand{\hau}{\mathcal{\hat A}_u}
\newcommand{\haun}{\mathcal{\hat A}_u^n}
\newcommand{\aud}{\mathcal{A}_{u,d}}
\newcommand{\haud}{\mathcal{\hat A}_{u,d}}
\newcommand{\haudp}{\mathcal{\hat A}'_{u,d}}
\newcommand{\hD}{\mathcal{\hat {\vec D}}}
\newcommand{\cdn}{\hat C_d^n}
\newcommand{\tr}{\mathrm{tr}}
\newcommand{\hj}{\hat J}
\newcommand{\hjd}{\hat J_d}

\newcommand{\hjud}{{\hat J}_{u,d}}
\newcommand{\hjq}{\hat J_Q}
\newcommand{\ltev}{\left( \frac{\Lambda_{\rm NP}}{1\,\textrm{TeV}} \right)}
\newcommand{\xqo}{X_Q^{\Delta F=1}}
\newcommand{\xqt}{X_Q^{\Delta F=2}}

\newcommand{\lmc}{\lambda_{\rm C}}
\newcommand{\eg} {{\it e.g}}

\begin{document}

\title{\bf Covariant Description of Flavor Conversion in the LHC Era}
\author{Oram Gedalia, Lorenzo Mannelli and Gilad Perez
\vspace{6pt} \\ \fontsize{12}{16}\selectfont\textit{Department
of Particle Physics and Astrophysics, Weizmann Institute of
Science,} \\ \fontsize{12}{16}\selectfont\textit{Rehovot 76100,
Israel}}
\date{}
\maketitle

\begin{abstract}
A simple covariant formalism to describe flavor and CP
violation in the left-handed quark sector in a model
independent way is provided. The introduction of a covariant
basis, which makes the standard model approximate symmetry
structure manifest, leads to a physical and transparent picture
of flavor conversion processes. Our method is particularly
useful to derive robust bounds on models with arbitrary
mechanisms of alignment. Known constraints on flavor violation
in the $K$ and $D$ systems are reproduced in a straightforward
manner. Assumptions-free limits, based on top flavor violation
at the LHC, are then obtained. In the absence of signal, with
100\,fb$^{-1}$ of data, the LHC will exclude weakly coupled
(strongly coupled) new physics up to a scale of 0.6\,TeV
(7.6\,TeV), while at present no general constraint can be set
related to $\Delta t=1$ processes. LHC data will constrain
$\Delta F=2$ contributions via same-sign tops signal, with a
model independent exclusion region of 0.08\,TeV (1.0\,TeV).
However, in this case, stronger bounds are found from the study
of CP violation in $D-\overline D$ mixing with a scale of
0.57\,TeV (7.2\,TeV). In addition, we apply our analysis to
models of supersymmetry and warped extra dimension. The minimal
flavor violation framework is also discussed, where the
formalism allows to distinguish between the linear and generic
non-linear limits within this class of models.
\end{abstract}

\section{Introduction}
The standard model (SM) has a unique way of incorporating CP
violation (CPV) and suppressing flavor changing neutral
currents (FCNCs). In fact, the way the SM flavor symmetry is
broken to allow flavor conversion is quite intriguing. It can
be described in the language of collective
breaking~\cite{Kagan:2009bn,Perez:2009xj}, a term commonly used
in the Little Higgs literature (see \eg~\cite{LHRev} and refs.
therein). Inter-generation transitions require the presence of
non-universal Yukawa couplings for both up and down quarks, a
non-vanishing weak coupling and a non-trivial CKM matrix. The
lightness of the first two generation masses and the
approximate alignment between the Yukawa matrices further
suppress FCNC transitions involving the first two generation.
This is manifest in particular in processes which are
characterized by hard GIM, such as ones involving CPV.
Inclusive third generation processes are further simplified due
to the presence of an approximate residual $U(1)_Q$
symmetry~\cite{Gedalia:2010zs}, only broken by the mass
differences of the light quarks.

New types of microscopic dynamics with a different flavor
breaking machinery typically give rise to deviation from the SM
approximated selection rules, and hence can be distinctively
distinguished from the SM. Till today no deviation from the SM
predictions related to quark flavor violation has been
observed\footnote{Recently, a hint for such a deviation has
been observed at the Tevatron in the like-sign dimuon charge
asymmetry~\cite{Abazov:2010hv}.}. This probably implies that
new physics (NP) searches should focus on SM extensions which,
if not flavor blind, share some of the structure and properties
described above.

Regarding the first two generations, models which do not
include some sort of degeneracies or flavor alignment (that is,
when NP contributions are diagonal in the quark mass basis) are
bounded to a high energy scale. Moreover, contributions
involving only quark doublets cannot be simultaneously aligned
with both the down and the up mass bases, hence even alignment
theories are constrained by measurements. However, the
hierarchy problem is not triggered by the light quarks, but
rather by the large top Yukawa, where almost any natural NP
model consists of an extended top sector. In addition, within
the SM, the top dominates the CP violating transitions, and
dials the amount of custodial symmetry breaking. Ironically,
the top sector is the least experimentally explored, and at
present model independent bounds on its flavor violating
couplings are rather poor.

In this work, we elaborate on a basis independent formalism for
studying flavor constraints in the quark sector, that was
recently introduced by us in~\cite{Gedalia:2010zs} (see
also~\cite{Feldmann:2009dc} for related work about algebraic
flavor invariants). Apart from yielding a simple, symmetry
driven, manner to understand the SM way of breaking flavor and
CP, it also provides a straightforward method to study generic
forms of NP flavor violation and derive model independent
bounds (focusing on the left-handed quark sector and assuming
$SU(2)_L$-invariant NP contributions). We start with a two
generations analysis, where a natural geometric interpretation
can be applied. It allows us to straightforwardly reproduce
known results~\cite{Blum:2009sk}. We then consider the three
generations case, where a dramatic improvement in the
measurements related to the top sector is expected at the LHC.
Thus, it is rather interesting to asses, before the data is
analyzed, what is the potential impact of the projected
sensitivity on beyond the SM searches. Our formalism makes
manifest the SM approximate $U(2)$ symmetry, due to the
lightness of the first two generation masses, for the up and
down quark sectors. In this limit, the SM actually posses a
residual $U(1)_Q$ symmetry, which is automatically incorporated
by our formalism. Under this symmetry, the massless first two
generations break into an ``active'' one, which interacts with
the heavy state, and a non-interactive ``sterile'' state. This
description is useful, not only conceptually, but also when
considering top and jet physics at the LHC, which in practice
cannot distinguish between light quark jets. The combination of
data from the down and the up sectors is used to robustly
constrain models including arbitrary mechanisms of alignment.

The analysis is based on the SM flavor group for quarks:
\beq
G_{SM}=U(3)_Q\times U(3)_U \times U(3)_D \, ,
\eeq
where $Q$, $U$ and $D$ stand for quarks doublets, up-type
singlets and down-type singlets, respectively. As mentioned,
$G_{SM}$ is broken within the SM only by the Yukawa
interactions. Therefore, we can treat the Yukawa matrices $Y_u$
and $Y_d$ as spurions, which transform as
$(\mathbf{3},\bar{\mathbf{3}},1)$ and
$(\mathbf{3},1,\bar{\mathbf{3}})$, respectively, under the
flavor group. In order to attain a covariant geometric picture,
we need to construct objects out of the Yukawa matrices which
transform in the same way. These are simply $Y_u Y_u^\dagger$
and $Y_d Y_d^\dagger$, which are in the $(\mathbf{8}+1,1,1)$
representation. Since the trace of these matrices does not
affect flavor changing processes, it is useful to remove it,
and work with $\yut$ and $\ydt$, both of which are adjoints of
$U(3)_Q$. For simplicity of notation, we denote these objects
as
\beq
\au \equiv \yut \, , \qquad \ad \equiv \ydt \, .
\eeq
As shown below, we can use these SM spurions of flavor
violation to construct a covariant basis. This basis turns out
to physically describe the flavor violation of the SM, as well
as of NP.

This paper is organized as follows: The two generations case,
for which a geometric formalism is devised, is discussed in
Sec.~\ref{sec:2g}. The covariant description for third
generation flavor violation is given in Sec.~\ref{sec:3g}. In
Sec.~\ref{sec:app} we use our formalism to constrain NP models
in an assumption-free manner, based on third generation $\Delta
F=1$ decays. Sec.~\ref{sec:uutt} similarly deals with $\Delta
F=2$ processes involving the third generation quarks. For the
latter two sections, current experimental data is used for the
down sector constraints, while the up sector bounds are mostly
based on LHC prospects. Secs.~\ref{sec:susy} and~\ref{sec:rs}
present concrete examples for the application of the analysis
to supersymmetry and warped extra dimension, respectively.
Finally, we conclude in Sec.~\ref{sec:conc}.

\section{Two Generations} \label{sec:2g}
We start with the simpler two generations case, which is
actually very useful in constraining new physics, as a result
of the richer experimental data. Any hermitian traceless $2
\times 2$ matrix can be expressed as a linear combination of
the Pauli matrices $\sigma_i$. This combination can be
naturally interpreted as a vector in three dimensional real
space, which applies to $\ad$ and $\au$. We can then define a
length of such a vector, a scalar product, a cross product and
an angle between two vectors, all of which are
basis-independent\footnote{The factor of $-i/2$ in the cross
product is required in order to have the standard geometrical
interpretation $\left| \vec{A} \times \vec{B} \right|
=|\vec{A}||\vec{B}|\sin \theta_{AB}$, with $\theta_{AB}$
defined through the scalar product as in
Eq.~\eqref{definitions}.}:
\beq \label{definitions}
\begin{split}
|&\vec{A}| \equiv \sqrt{\frac{1}{2} \tr(A^2)} \, , \quad
\vec{A} \cdot \vec{B} \equiv \frac{1}{2} \tr(A \, B) \, , \quad
\vec{A}
\times \vec{B} \equiv -\frac{i}{2} \left[ A,B \right] \, , \\
&\cos (\theta_{AB}) \equiv \frac{\vec{A} \cdot
\vec{B}}{|\vec{A}| |\vec{B}|}= \frac{\tr(A \, B)}{\sqrt{\tr
(A^2) \tr (B^2)}} \, .
\end{split}
\eeq

\begin{figure}[hbt]
\centering
\includegraphics[width=2.2In]{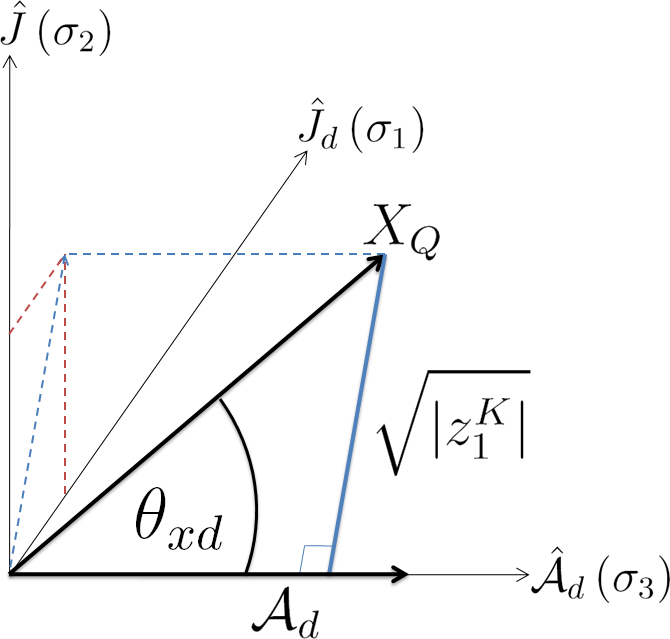}
\caption{The contribution of $X_Q$ to $K^0-\overline{K^0}$ mixing, $\Delta m_K$, given by
the solid blue line. In the down mass basis, $\had$ corresponds to $\sigma_3$,
$\hj$ is $\sigma_2$ and $\hjd$ is $\sigma_1$.}
\label{fig:2g_fv}
\end{figure}

These definitions allow for an intuitive understanding of the
flavor and CP violation induced by a new physics source.
Consider a dimension six $SU(2)_L$-invariant operator,
involving only quark doublets,
\beq \label{o1}
\frac{z_1}{\Lambda_{\rm NP}^2} O_1=\frac{1}{\Lambda_{\rm NP}^2}
\left( \overline{Q}_{i} (X_Q)_{ij} \gamma_\mu Q_{j} \right)
\left( \overline{Q}_{i} (X_Q)_{ij} \gamma^\mu Q_{j} \right) \,
,
\eeq
where $\Lambda_{\rm NP}$ is some high energy scale and $z_1$ is
the Wilson coefficient. $X_Q$ is a traceless hermitian matrix,
transforming as an adjoint of $SU(3)_Q$ (or $SU(2)_Q$ for two
generations), so it ``lives'' in the same space as $\ad$ and
$\au$.\footnote{This operator can always be written as a
product of two identical adjoints, as explained in
Appendix~\ref{app:adj}.} In the down sector for example, the
operator above is relevant for flavor violation through
$K^0-\overline{K^0}$ mixing. To analyze its contribution, we
define a covariant basis for each sector, with the following
unit vectors
\beq \label{2g_basis}
\haud \equiv \frac{\aud}{\left| \aud \right|} \, , \quad \hj
\equiv \frac{\ad \times \au}{\left|\ad \times \au\right|} \, ,
\quad \hjud \equiv \haud \times \hj \, .
\eeq
Then the contribution of the operator in Eq.~\eqref{o1} to
$\Delta c,s=2$ processes is given by the misalignment between
$X_Q$ and $\aud$, which is equal to
\beq \label{2g_fv}
\left| z_1^{D,K} \right|=\left| { X_Q} \times {\haud} \right|^2
\, .
\eeq
This result is manifestly invariant under a change of basis.
The meaning of Eq.~\eqref{2g_fv} can be understood as follows:
We can choose an explicit basis, for example the down mass
basis, where $\ad$ is proportional to $\sigma_3$. $\Delta s=2$
transitions are induced by the off-diagonal element of $X_Q$,
so that $\left| z_1^K \right|=\left| (X_Q)_{12} \right|^2$.
Furthermore, $\left| (X_Q)_{12} \right|$ is simply the combined
size of the $\sigma_1$ and $\sigma_2$ components of $X_Q$. Its
size is given by the length of $X_Q$ times the sine of the
angle between $X_Q$ and $\ad$ (see Fig~\ref{fig:2g_fv}). This
is exactly what Eq.~\eqref{2g_fv} describes.

\begin{figure}[htb]
\centering
\includegraphics[width=2.7In]{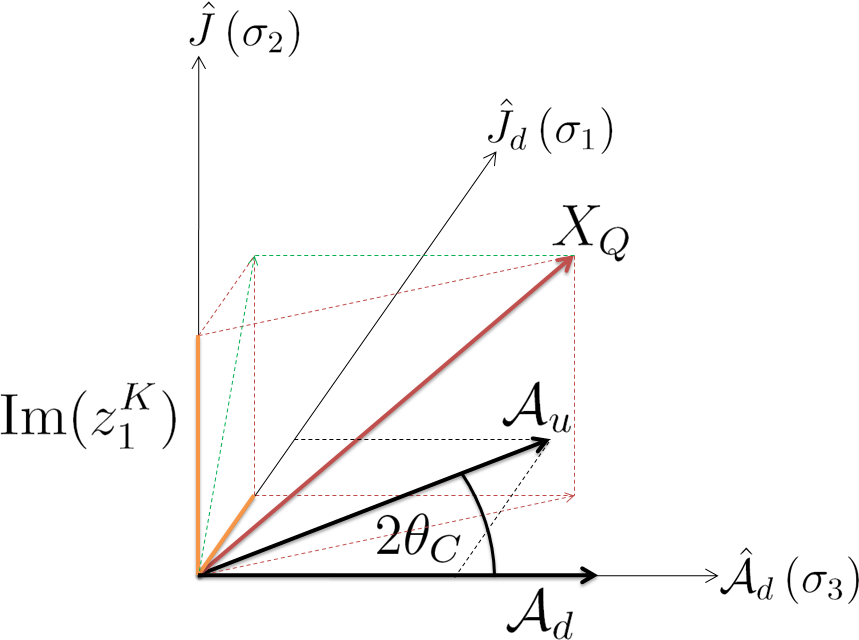}
\caption{CP violation in the Kaon system induced by $X_Q$. $\mathrm{Im}(z_1^K)$ is twice
the product of the two solid orange lines, which are the projections of $X_Q$ on the $\hj$
and $\hjd$ axes. Note that the angle between $\ad$ and $\au$ is twice the Cabibbo angle, $\theta_C$.}
\label{fig:2g_cp}
\end{figure}

Next we discuss CPV, which is given by
\beq \label{cpv_2g}
\mathrm{Im}\left(z_1^{K,D}\right)=2\left(X_Q \cdot \hat
J\right)\left( X_Q \cdot \hat J_{u,d} \right)\,.
\eeq
The above expression is easy to understand in the down basis,
for instance. In addition to diagonalizing $\ad$, we can also
choose $\au$ to reside in the $\sigma_1-\sigma_3$ plane
(Fig.~\ref{fig:2g_cp}) without loss of generality, since there
is no CPV in the SM for two generations. As a result, all of
the potential CPV originates from $X_Q$ in this basis. $z_1^K$
is the square of the off-diagonal element in $X_Q$,
$(X_Q)_{12}$, thus Im$\left(z_1^K\right)$ is simply twice the
real part ($\sigma_1$ component) times the imaginary part
($\sigma_2$ component). In this basis we have $\hj\propto
\sigma_1$ and $\hjd\propto \sigma_2$, this proves the validity of Eq.~\eqref{cpv_2g}.

The weakest unavoidable bound coming from measurements in the
$K$ and $D$ systems was derived in~\cite{Blum:2009sk} using
a specific parameterization of $X_Q$. In the covariant bases defined
in Eq.~\eqref{2g_basis}, $X_Q$ can be written as
\beq \label{xq_2g}
X_Q=X^{u,d} \haud+X^J \hj+X^{J_{u,d}} \hjud \,,
\eeq
and the two bases are related through
\beq \label{u_to_d_2g}
X^u=\cos 2\theta_{\rm C} X^d-\sin 2\theta_{\rm C} X^{J_d} \, ,
\quad X^{J_u}=-\sin 2\theta_{\rm C} X^d-\cos 2\theta_{\rm C}
X^{J_d} \, ,
\eeq
while $X^J$ remains invariant. Plugging Eqs.~\eqref{xq_2g}
and~\eqref{u_to_d_2g} into Eqs.~\eqref{2g_fv}
and~\eqref{cpv_2g}, we obtain explicit results. It is then easy
to see that in the parameterization employed
in~\cite{Blum:2009sk}, $\Lambda_{12} \sin \gamma$ is equal to
$X^J$, $\Lambda_{12} \sin \alpha \cos \gamma$ is equal to
$X^{J_d}$ etc., therefore their results coincide with ours.

An interesting conclusion can be inferred from the analysis
above: In addition to the known necessary condition for CPV in
two generation~\cite{Blum:2009sk}
\beq \label{cpv_cond}
X^J \propto \tr \left( X_Q \left[ \ad,\au \right] \right) \neq
0 \,,
\eeq
we identify a second necessary condition, exclusive for $\Delta
F=2$ processes:
\beq \label{cpv_new_cond}
X^{J_{u,d}} \propto\tr \left( X_Q \left[ \aud , \left[ \ad,\au
\right] \right] \right) \neq 0 \, ,
\eeq
The strength of these conditions is that they involve only the
basic physical ingredients $\aud$ and $X_Q\,$, and they can be
clearly identified from the geometric interpretation. Note,
however, that this new condition in Eq.~\eqref{cpv_new_cond} is
only applicable to either the down or the up sector, while the
known condition in Eq.~\eqref{cpv_cond} is universal.

\section{Three Generations} \label{sec:3g}
\subsection{Approximate $U(2)$ Limit of Massless Light Quarks}
\label{sec:3g_u2}
\begin{figure}[htb]
\centering
\includegraphics[width=2.33In]{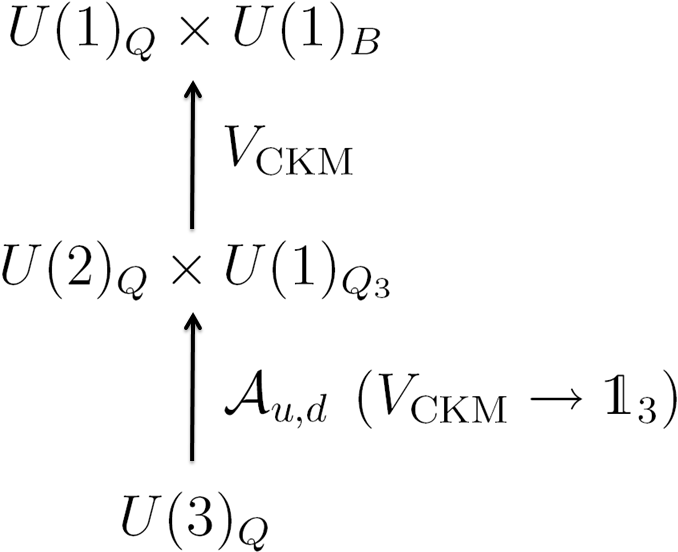}
\caption{The flavor symmetry breaking pattern for the left-handed sector:
 The $U(3)_Q$ group is broken by each of $\au$ and $\ad$ to an approximate
  $U(2)_Q \times U(1)_{Q_3}$; including the fact that these two objects are
  not aligned (that is, the CKM matrix is not trivial), the symmetry is
  broken into baryon number and an approximate $U(1)_Q$ for the light quarks
   combination that effectively decouples.}
\label{fig:breaking}
\end{figure}

For three generations, a simple 3D geometric
interpretation does not naturally emerge anymore, as the
relevant space is characterized by the eight Gell-Mann
matrices\footnote{We denote the Gell-Mann matrices by
$\Lambda_i$, where $\tr(\Lambda_i \Lambda_j)=2\delta_{ij}$.
Choosing this convention allows us to keep the definitions of
Eq.~\eqref{definitions}.}. A useful approximation appropriate
for third generation flavor violation is to neglect the masses
of the first two generation quarks, where the breaking of the
flavor symmetry is characterized by
$[U(3)/U(2)]^2$~\cite{Kagan:2009bn}. This description is especially suitable
for the LHC, where it would be difficult to distinguish between
light quark jets of different flavor. In this limit, the 1-2
rotation and the phase of the CKM matrix become unphysical, and
we can, for instance, further apply a $U(2)$ rotation to the first two
generations to ``undo'' the 1-3 rotation.
Therefore, the CKM matrix is effectively reduced to a real
matrix with a single rotation angle between an active light
flavor (say, the 2nd one) and the 3rd generation,
\beq \label{theta}
\theta\cong \sqrt{\theta_{13}^2+\theta_{23}^2}\,,
\eeq
where $\theta_{13}$ and $\theta_{23}$ are the corresponding CKM
mixing angles. The other generation (the first one) decouples,
and is protected by a residual $U(1)_Q$ symmetry. This can be
easily seen when writing $\ad$ and $\au$ in, say, the down mass
basis
\beq \label{3g_yukawas}
\ad = \frac{y_b^2}{3} \begin{pmatrix} -1 & 0 & 0\\ 0 & -1 & 0\\
0 & 0 & 2 \end{pmatrix} \, , \qquad \au= y_t^2
\begin{pmatrix} \spadesuit & 0 & 0 \\ 0 & \spadesuit & \spadesuit \\
0 & \spadesuit & \spadesuit \end{pmatrix} \, ,
\eeq
where $\spadesuit$ stands for a non-zero \emph{real} entry. The
resulting flavor symmetry breaking scheme is depicted in
Fig.~\ref{fig:breaking}.

An interesting consequence of this approximation is that a
complete basis cannot be defined covariantly, since $\aud$ in
Eq.~\eqref{3g_yukawas} clearly span only a part of the eight
dimensional space. More concretely, we can identify four
directions in this space: $\hj$ and $\hjud$ from
Eq.~\eqref{2g_basis} and either one of the two orthogonal pairs
\beq \label{3g_basis1}
\haud \quad \mathrm{and} \quad \hat C_{u,d} \equiv 2 \hj \times
\hjud-\sqrt{3} \haud \, ,
\eeq
or
\beq \label{3g_basis2}
\haudp \equiv \hj \times \hjud \quad \mathrm{and} \quad \hjq
\equiv \sqrt3 \hj \times \hjud-2 \haud \, .
\eeq
Note that $\hjq$ corresponds to the conserved $U(1)_Q$
generator, so it commutes with both $\ad$ and $\au$, and takes
the same form in both bases\footnote{The meaning of these basis
elements can be understood from the following: In the down mass
basis we have $\had=-\Lambda_8$, $\hj=\Lambda_7$,
$\hjd=\Lambda_6$ and $\hat C_d=\Lambda_3$. The alternative
diagonal generators from Eq.~\eqref{3g_basis2} are
$\hadp=(\Lambda_3-\sqrt3 \Lambda_8)/2=\mathrm{diag}(0,-1,1)$
and $\hjq=(\sqrt3 \Lambda_3+\Lambda_8)/2=
\mathrm{diag}(2,-1,-1)/\sqrt3$. It is then easy to see that
$\hjq$ commutes with the effective CKM matrix, which is just a
2-3 rotation, and that it corresponds to the $U(1)_Q$
generator, $\mathrm{diag}(1,0,0)$, after trace subtraction and
proper normalization.}. There are four additional directions,
collectively denoted as $\hD$, which transform as a doublet
under the CKM (2-3) rotation, and do not mix with the other
generators. The fact that these cannot be written as
combinations of $\aud$ stems from the approximation introduced
above of neglecting light quark masses. Without this
assumption, it is possible to span the entire space using the
Yukawa matrices~\cite{Colangelo:2008qp,{Mercolli:2009ns},
Ellis:2009di}. Despite the fact that this can be done in several ways, in the next subsection we focus on a realization for which the basis elements have a clear physical meaning.

\begin{figure}[htb]
\centering
\includegraphics[width=2.5In]{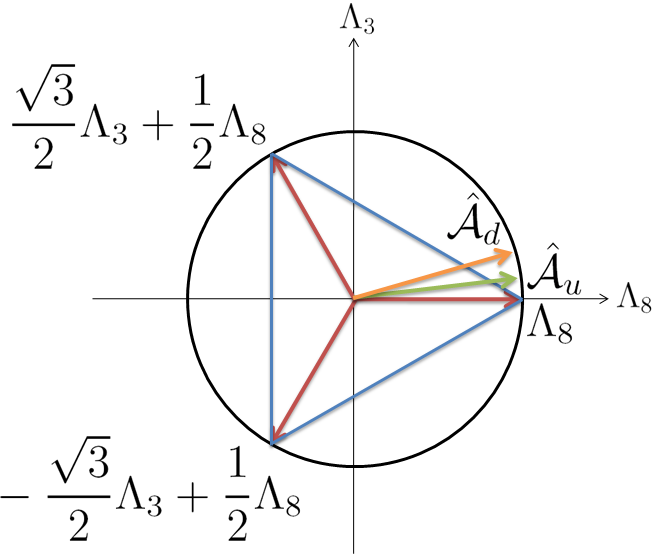}
\caption{The three unit-length diagonal traceless matrices with an inherent
$SU(2)$ symmetry. $\had$ and $\hau$ were schematically added
(their angle to the $\Lambda_8$ axis is actually much smaller than what appears in the plot).}
\label{fig:u2}
\end{figure}

It is interesting to notice that a given traceless adjoint
object $X$ in three generations flavor space has an inherent
$SU(2)$ symmetry (that is, two identical eigenvalues) if and
only if it satisfies
\beq
\left[\tr \left( X^2\right)\right]^{3/2}=\sqrt6\, \tr \left(
X^3 \right) \, .
\eeq
In this case it must be a unitary rotation of either
$\Lambda_8$ or its permutations $(\Lambda_8 \pm \sqrt3
\Lambda_3)/2$, which form an equilateral triangle in the
$\Lambda_3-\Lambda_8$ plane (see Fig.~\eqref{fig:u2}).

As before, we wish to characterize the flavor violation induced
by $X_Q$ in a basis independent form. The simplest observable
we can construct is the overall flavor violation of the third
generation quark, that is, its decay to any quark of the first
two generations. This can be written as
\beq \label{inclusive_decay}
\frac{2}{\sqrt{3}} \left| X_Q \times \haud \right| \, ,
\eeq
which extracts $\sqrt{\left| (X_Q)_{13}\right|^2+\left|
(X_Q)_{23}\right|^2}$ in each basis.

\subsection{No $U(2)$ Limit~-- Complete Covariant Basis}
\label{sec:3g_full}

It is sufficient to restore the masses of the second generation
quarks in order to describe the full flavor space. A
simplifying step to accomplish this is to define the following object: We take the
$n$-th power of $\left( Y_d Y_d^\dagger \right)$, remove the
trace, normalize and take the limit $n \to \infty$. This is
denoted by $\hadn$:
\beq
\hadn \equiv \lim_{n \to \infty} \left\{ \frac{\left( Y_d
Y_d^\dagger\right)^n-\mathds{1}_3\tr \left[ \left( Y_d
Y_d^\dagger\right)^n \right]/3}{\left| \left( Y_d
Y_d^\dagger\right)^n-\mathds{1}_3\tr \left[ \left( Y_d
Y_d^\dagger\right)^n \right]/3 \right|}\right\} \,,
\eeq
and we similarly define $\haun$.
Once we take the limit $n\to \infty$ the small eigenvalues of $\haud$ go to zero
and the approximation assumed before is formally
reproduced. As before, we compose the following basis elements:
\beq
\hj^n \equiv \frac{\hadn \times \haun}{\left| \hadn \times
\haun \right|} \, , \quad \hjd^n \equiv \frac{\hadn \times
\hj^n}{\left| \hadn \times \hj^n \right|} \, , \quad \cdn
\equiv 2 \hj^n \times \hjd^n -\sqrt{3} \had^n \, ,
\eeq
which are again identical to the previous case. The important
observation for this case is that the $U(1)_Q$ symmetry is now
broken. Consequently, the $U(1)_Q$ generator, $J_Q$, does not
commute with $\ad$ and $\au$ anymore (nor does $\cdn$, which is
different from $J_Q$ only by normalization and a shift by
$\ad$, see Eqs.~\eqref{3g_basis1} and~\eqref{3g_basis2}). It is
thus expected that the commutation relation $[ \ad,\cdn]$
(where $\ad$ now contains also the strange quark mass) would
point to a new direction, which could not be obtained in the
approximation used before. Further commutations with the
existing basis elements should complete the description of the
flavor space.

We thus define
\beq
\hat D_2 \equiv \frac{\had \times \cdn}{\left| \had \times \cdn
\right|} \, .
\eeq
In order to understand the physical interpretation, note that
$\hat D_2$ does not commute with $\ad$, so it must induce
flavor violation, yet it does commute with $\hadn$. The latter
can be identified as a generator of a $U(1)$ symmetry for the
bottom quark (it is proportional to diag(0,0,1) in its diagonal
form, without removing the trace), so this fact means that
$\hat D_2$ preserves this symmetry. Therefore it must represent
a transition between the first two generations of the down
sector.

We further define
\beq
\hat D_1 \equiv \frac{\had \times \hat D_2}{\left| \had \times
\hat D_2 \right|} \, , \quad \hat D_4 \equiv \frac{\hjd^n
\times \hat D_2}{\left| \hjd^n \times \hat D_2 \right|} \, ,
\quad \hat D_5 \equiv \frac{\hj^n \times \hat D_2}{\left| \hj^n
\times \hat D_2 \right|} \, ,
\eeq
which complete the basis. All of these do not commute with
$\ad$, thus producing down flavor violation. $\hat D_1$
commutes with $\hadn$, so it is of the same status as $\hat
D_2$. The last two elements, $\hat D_{4,5}$, are responsible
for third generation decays, similarly to $\hj^n$ and $\hjd^n$.
More concretely, the latter two involve transitions between the
third generation and what was previously referred to as the
``active'' generation (a linear combination of the first two),
while $\hat D_{4,5}$ mediate transitions to the orthogonal
combination. It is of course possible to define linear
combinations of these four basis elements, such that the decays
to the strange and the down mass eigenstates are separated, but
we do not proceed with this derivation. It is also important to
note that this basis is not completely orthogonal. An explicit
decomposition of all the covariant objects in a specific basis
can be found in Appendix~\ref{app:decomp}.

An instructive exercise is to decompose $\au$ in this covariant
``down'' basis, since $\au$ is a flavor violating source within
the SM. Focusing only on the dependence on the small parameters
$\lambda_{\rm C}$ and $y_c^2/y_t^2$ (and omitting for
simplicity $\mathcal{O}(1)$ factors such as the Wolfenstein
parameter $A$), we have
\beq \label{au_cov_decomp}
\begin{split}
\au& \cdot \left\{\hat D_1,\hat D_2,\cdn,\hat D_4,\hat
D_5,\hjd^n,\hj^n,\hadn\right\} \sim \\ &\left\{\lmc y_c^2+
\lmc^5 y_t^2,\lmc y_c^2,(y_c^2+ \lmc^4 y_t^2)/2,\lmc^3
y_c^2,\lmc^3 y_c^2,\lmc^2 y_t^2,0,y_t^2/\sqrt3 \right\} \,.
\end{split}
\eeq
This shows the different types of flavor violation in the down
sector within the SM. It should be mentioned that the $\hat
D_2$ and $\hat D_5$ projections of $\au$ vanish when the CKM
phase is taken to zero, and also when either of the CKM mixing
angles is zero or $\pi/2$. Therefore these basis elements can
be interpreted as CP violating, together with $\hj^n$. As an
example, notice that a $2 \to 1$ transition in the down sector,
represented by the projection to $\hat D_1$, can either occur
via mixing with the third generation at the order $\lmc^5
y_t^2$ or among the first two generations only at the order
$\lmc y_c^2\,$. Yet CPV in this transition can only be
generated through the latter type of contribution at $\lmc
y_c^2\,$, as can be seen from the $\hat D_2$ projection (recall
again that these are not the $d$ and $s$ mass eigenstates, but
instead the ``active'' and ``inactive'' generations, after a
$U(2)$ rotation has been applied). Analogously, a $3 \to 1$
transition occurs at $\lmc^3 y_c^2$ whether it is CP conserving
or CP violating, as inferred from the $\hat D_{4,5}$
projections.

In the rest of the paper we use the description based on the
approximate $U(2)$ symmetry, rather than the full basis,
whenever possible.

\section{Third Generation $\Delta F=1$ Transitions} \label{sec:app}
We now use measurements from the down and the up sectors to
derive a model independent bound on the corresponding NP scale,
based on the overall flavor violating decay of the third
generation quarks. We focus on the following operator
\beq \label{3g_operator}
O^h_{LL}= i \left[ \overline{Q}_i \gamma^\mu (\xqo)_{ij} Q_j
\right] \left[ H^\dagger \overleftrightarrow{D}_\mu H \right] +
\mathrm{h.c.} \, ,
\eeq
which contributes at tree level to both top and bottom
decays~\cite{Fox:2007in}\footnote{It is important to note that
a given NP model might generate different higher-dimensional
operators via different types of processes (the general from of
the relevant low energy effective theory is discussed for
example in~\cite{tEFT}). Therefore $X_Q$ is in general
different for each operator, so we denote it specifically as
$\xqo$ for the current case.}. Note that as in the two
generations case, we only deal with the left-handed sector,
where down and up contributions are related. We omit an
additional operator for quark doublets, $O^u_{LL}= i\left[
{\overline Q}_3 {\tilde H} \right] \left[ \big( D\!\!\!\!\slash
{\tilde H} \big)^\dagger Q_2 \right] - i\left[ {\overline Q}_3
\big( D\!\!\!\!\slash {\tilde H} \big) \right] \left[ {\tilde
H}^\dagger Q_2 \right]$, which induces bottom decays only at
one loop, but in principle it should be included in a more
detailed analysis.

The experimental constraints we use
are~\cite{Aubert:2004it,Iwasaki:2005sy,Carvalho:2007yi}
\beq \label{tbdecay}
\begin{split}
\mathrm{Br}&(B \to X_s\ell^+ \ell^-)_{1 \textrm{
GeV}^2<q^2<6 \textrm{ GeV}^2}=(1.61 \pm 0.51) \times 10^{-6} \, , \\
\mathrm{Br}&(t\to (c,u)Z) <5.5 \times 10^{-5} \, ,
\end{split}
\eeq
where the latter corresponds to the prospect of the LHC bound
in the absence of signal for 100~fb$^{-1}$. We adopt the
weakest limits on the coefficient of the operator in
Eq.~\eqref{3g_operator}, $C^h_{LL}$, derived
in~\cite{Fox:2007in}:
\beq \label{exp_constraints}
\begin{split}
\mathrm{Br}&(B \to X_s\ell^+ \ell^-) \longrightarrow \left|
C^h_{LL} \right|_b < 0.018 \ltev^2 \, , \\ \mathrm{Br}&(t\to
(c,u)Z) \longrightarrow \left| C^h_{LL} \right|_t < 0.18
\ltev^2 \, ,
\end{split}
\eeq
and define $r_{tb} \equiv \left| C^h_{LL} \right|_t/\left|
C^h_{LL} \right|_b\,$.

The NP contribution can be decomposed in the covariant bases as
\beq \label{xq_param}
\xqo = X'^{u,d} \haudp +X^J \hj+X^{J_{u,d}} \hjud +X^{J_Q} \hjq
+ X^{\vec D} \hD \,.
\eeq
The length of $\xqo$ is denoted, based on the definition in
Eq.~\eqref{definitions}, by
\beq
L \equiv \left| \xqo \right| \, .
\eeq
The weakest bound is obtained, for a fixed $L$, by finding a
direction of $X_Q$ that minimizes the contributions to $\left|
C^h_{LL} \right|_t$ and $\left| C^h_{LL} \right|_b$, thus
constituting the ``best'' alignment. However, since $\hjq$
commutes with $\aud$, as discussed above, it does not
contribute to third generation decay
(Eq.~\eqref{inclusive_decay}) in neither sectors. On the other
hand, any component of $\xqo$ may also generate flavor
violation among the first two generations (when their masses
are switched back on), which is more strongly constrained.
Specifically, the bound that stems from the case of $\xqo
\propto \hjq$, derived in Appendix~\ref{app}, is
\beq \label{3g2g_constraint}
L<0.59 \ltev^2; \quad  \Lambda_{\rm NP}>1.7 \, \mathrm{TeV} \,
,
\eeq
where the latter is for $L=1$. This is stronger than the limit
given below for other forms of $\xqo$, hence this does not
constitute the optimal alignment. To conclude this issue, all
directions that contribute to first two generations flavor and
CPV at ${\cal O}\left(\lambda_{\rm C}\right)$, that is $\hjq$,
$\hD$ and $\haudp$, are not favorable in terms of alignment, as
discussed in Appendix~\ref{app}.

The induced third generation flavor violation, after removing
these contributions, is then given by
\beq \label{3gfv_explicit}
\frac{4}{3} \left| \xqo\times \haud \right|^2 =
\left(X^J\right)^2+ \left(X^{J_{u,d}}\right)^2  \,,
\eeq
and in order to see this in a common basis, we express
$X^{J_u}$ as
\beq \label{xju}
X^{J_u}=\cos 2\theta \,X^{J_d}+ \sin 2\theta \,X'^d\,,
\eeq
with $\theta$ as defined in Eq.~\eqref{theta}. From this it is
clear that $X^J$ contributes the same to both the top and the
bottom decay rates, so it should be set to zero for optimal
alignment. Thus the best alignment is obtained by varying
$\alpha$, defined by
\beq \label{tanalpha}
\tan\alpha \equiv \frac{X^{J_{d}}}{X^d}\,.
\eeq
Here we use $X^d$, which is the coefficient of $\had$, instead
of $X'^d$, since the former does not produce flavor violation
among the first two generations to leading order (up to
$\mathcal{O}(\lambda_{\rm C}^5)$).

\begin{figure}[hbt]
\centering
\includegraphics[width=4In]{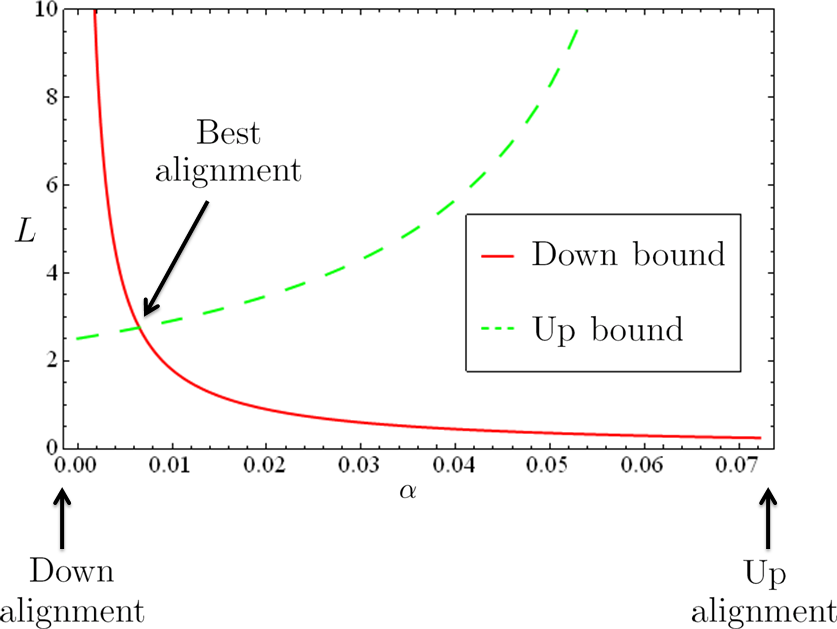}
\caption{Upper bounds on $L$ as a function of $\alpha$, coming from the measurements
of flavor violating decays of the bottom and the top quarks, assuming $\Lambda_{\rm NP}=1$ TeV.}
\label{fig:3g_bounds}
\end{figure}

We now consider two possibilities: (i) complete alignment with
the down sector; (ii) the best alignment satisfying the bounds
of Eq.~\eqref{exp_constraints}, which gives the weakest
unavoidable limit. Note that we can also consider up alignment,
but it would give a stronger bound than down alignment, as a
result of the stronger experimental constraints. The bounds for
these cases are~\cite{Gedalia:2010zs}
\beq \label{3g_bounds}
\begin{split}
\mathrm{(i)} & \quad \alpha=0 \, , \quad L<2.5 \ltev^2 ; \quad
\Lambda_{NP}>0.63 \,(7.9)\; \mathrm{TeV} \, ,
\\ \mathrm{(ii)} & \quad \alpha=\frac{\sqrt{3}\,\theta}{1+r_{tb}}
\, , \quad L<2.8 \ltev^2 ; \quad \Lambda_{NP}>0.6 \, (7.6) \;
\mathrm{TeV} \, ,
\end{split}
\eeq
as shown in Fig.~\ref{fig:3g_bounds}, where in parentheses we
give the strong coupling bound, in which the coefficient of the
operators in Eqs.~\eqref{o1} and~\eqref{3g_operator} is assumed
to be $16 \pi^2$. Note that these are weaker than the bound in
Eq.~\eqref{3g2g_constraint}.

It is important to mention that the optimized form of $\xqo$
generates also $c \to u$ decay at higher order in $\lambda_{\rm
C}$, which might yield stronger constraints than the top decay.
In Appendix~\ref{app} it is shown that the resulting bound from
the former is actually much weaker than the one from the top.
Therefore, the LHC is indeed expected to strengthen the model
independent constraints.

\section{Third Generation $\Delta F=2$ Transitions} \label{sec:uutt}
In the previous section we used $\Delta F=1$ decays of third
generation quarks to obtain the weakest model independent
constraint. Here we do the same using $\Delta F=2$ processes.
For simplicity, we only consider complete alignment with
the down sector
\beq \label{xq2}
\xqt=L \had \,,
\eeq
as the constraints from this sector are much stronger. This
generates in the up sector top flavor violation, and also
$D^0-\overline{D^0}$ mixing at higher order. Yet there is no
top meson, as the top quark decays too rapidly to hadronize.
Instead, we analyze the process $pp \to tt$ (related to mixing
by crossing symmetry), which is most appropriate for the LHC.
It should be emphasized, however, that in this case the parton
distribution functions of the proton strongly break the
approximate $U(2)$ symmetry of the first two generations. The
simple covariant basis introduced in Sec.~\ref{sec:3g_u2},
which is based on this approximate symmetry, cannot be used as
a result. Furthermore, this LHC process is dominated by $uu \to
tt$, so we focus only on the operator involving up (and not
charm) quarks. We verified numerically that indeed the charm
contribution to this process is smaller by an order of
magnitude.

The production of same-sign tops was studied in the literature
in the context of different models (see
\textit{e.g.}~\cite{Larios:2003jq,Kraml:2005kb,Gao:2008vv} and
refs.~therein). The simplest way observe it at the LHC and
distinguish it from $t \bar{t}$ production, is based on the
dilepton mode, in which two same-sign (mostly positive sign)
leptons are produced from the top quarks. However, the
branching ratio of this mode is only about 5\%, and there are
several types of SM backgrounds, such as $W^+W^+qq\,$. We
therefore choose to adopt a realistic assumption of 1\%
efficiency for detecting same-sign tops at the
LHC~\cite{Larios:2003jq}, including b-tagging efficiency and
the necessary cuts to isolate the signal\footnote{Examples for
possible cuts are requiring some minimal invariant mass for one
or two pairings of a lepton and a b-tagged jet and a minimal
transverse momentum for the latter jets. The chosen cuts
strongly affect the efficiency~-- in~\cite{Gao:2008vv}, \eg,
they eliminate the background almost completely, but at the
cost of reducing the signal cross section to less than 0.2\% of
its original value. A detailed analysis of this issue, which
can be found in the literature, is out of the scope of the
paper.}. In any case, our conclusions are only mildly sensitive
to this assumption, as explained below.

In order to estimate the prospect for the LHC bound on
same-sign tops production, we calculated the $uu \to tt$ cross
section using MadGraph/MadEvent~\cite{Alwall:2007st}, as a $t$
(or $u$) channel process mediated by a heavy vector boson, the
mass of which is identified with $\Lambda_{\rm NP}$. The
resulting cross section for the LHC with center of mass energy
of 14, 10 and 7 TeV, and for the Tevatron, is given by
\beq
\sigma^{tt}= \left\{60,30,13,0.013 \right\} \left(
\frac{1\,\textrm{TeV}}{\Lambda_{\rm NP}} \right)^4 \,
\mathrm{pb}
\eeq
respectively\footnote{The simulation was actually performed
with a high mass for the new vector boson, to avoid producing
it on-shell. The result was then scaled down to 1 TeV as
$\Lambda_{\rm NP}^4$ (we also verified within the simulation
that this scaling is correct).}. This was matched onto the
operator in Eq.~\eqref{o1}. We then used the fact that the
cross section times the integrated luminosity must be lower
than 3 for a 95\% exclusion, in the absence of
signal~\cite{PDG}. Adding the assumption of 1\% signal
efficiency, we find
\beq
z_1^{tt}<7.1 \times 10^{-3} \ltev ^2 \, ,
\eeq
for 100 fb$^{-1}$ at a center of mass energy of 14 TeV. The
experimental constraint from CPV in the $D$ system
is~\cite{Gedalia:2009kh,otherD}
\beq
\mathrm{Im}(z_1^D)<1.1 \times 10^{-7} \ltev^2 \, ,
\eeq

The contribution of $\xqt$ to these processes is calculated by
applying a CKM rotation to Eq.~\eqref{xq2}. CPV in the $D$
system is then given by $\mathrm{Im}\left[ \left(\xqt
\right)^2_{12} \right]$, and $\left| \left( \xqt \right)_{13}
\right|^2$ describes $uu \to tt$. Note that we have
\beq
\begin{split}
\left(\xqt \right)_{12} &\cong -\sqrt3 \, L V_{ub} V^*_{cb} \,
,
\\ \left( \xqt \right)_{13} &\cong -\sqrt3 \, L V_{ub} V^*_{tb} \, ,
\end{split}
\eeq
with $V_{ij}$ as the CKM matrix elements. The resulting bounds
are
\beq \label{uutt_bound}
L<12 \ltev\,; \quad \Lambda_{\rm NP}>0.08 \, (1)\, \mathrm{TeV}
\, ,
\eeq
for $uu \to tt$ and
\beq \label{dd_bound}
L<1.8 \ltev\,; \quad \Lambda_{\rm NP}>0.57 \, (7.2)\,
\mathrm{TeV} \, ,
\eeq
for $D$ mixing. It should be mentioned that the bound in
Eq.~\eqref{uutt_bound} depends on the quartic root of the cross
section that was evaluated above, thus it is only mildly
sensitive to that calculation and to the efficiency assumption.
Interestingly, the bound that stems from the Tevatron with 5
fb$^{-1}$ (assuming that same-sign top pairs were searched for
and not detected) is weaker than Eq.~\eqref{uutt_bound} by a
factor of $\sim$17.

The limits in Eqs.~\eqref{uutt_bound} and~\eqref{dd_bound} can
be further weakened by optimizing the alignment between the
down and the up sectors, as in the previous section. Since this
would only yield a marginal improvement of about 10\%, we do
not analyze this case in detail.

To conclude, we learn that for $\Delta F=2$ processes, the
existing bound is stronger than the one which will be obtained
at the LHC for top quarks, as opposed to $\Delta F=1$ case
considered above.

\section{Supersymmetry} \label{sec:susy}
The analysis presented above uses a model-independent language
via effective field theory. Here we apply our results to two SM
extensions~-- supersymmetry (SUSY) and the Randall-Sundrum (RS)
model of a warped extra dimension (in the next section).

We focus now on both $\Delta F=1$ and $\Delta F=2$ left-handed
processes within supersymmetric extensions of the SM. The idea
as in the above is to provide robust bounds, which could be
applied even to SUSY alignment models~\cite{Nir:1993mx} (for a
possible connection with bounds from EDM
see~\cite{Altmannshofer:2010ad}). The analysis of $\Delta F=1$
transitions is more involved as follows. The relevant
contributions to the left-handed operators is driven by the
squark doublets mass matrix, which transforms as an adjoint of
the minimal supersymmetric standard model (MSSM) flavor group.
However, the contributions to top and bottom decays are induced
by different operators in the effective Hamiltonian, hence our
treatment above does not apply. Instead we rederive the
relevant bounds on the squark mass matrix explicitly.

Given the large number of parameters involved in flavor
changing processes, it is often convenient to use the mass
insertion (MI) formalism. The mass insertions are defined in
the so-called super CKM basis. In this basis all the neutral
gaugino couplings $\tilde{g},\tilde{\gamma},\tilde{Z}$ are
flavor diagonal, and the charged $\tilde{W}^{\pm}$ quark-squark
mixing angles are equal to the CKM angles. In general, the
squarks mass matrices $\tilde m^{u,d}$ are not diagonal in the
Super CKM basis. Flavor violation is induced by the $\tilde
m^{u,d}$ off diagonal elements, and can be parameterized in
terms of the ratios
\beq
\left(\delta_{ij}^{f}\right)_{AB} = \frac{\left(\tilde
m_{ij}^{f}\right)_{AB}^{2}}{\tilde{m}_{Q}^{2}} \,,
\eeq
where $\left(\tilde m_{ij}^{f}\right)_{AB}^{2}$ are the
off-diagonal elements of the $f=u,d$ mass squared matrix that
mixes flavors $i,j$ for both left- and right-handed scalars
($A,B=\mbox{Left, Right}$), and where $\tilde{m}_{Q}$ indicates
the average squark mass.

\subsection{Top Decay}

\begin{figure}[h]
\centering{}\includegraphics[scale=0.6]{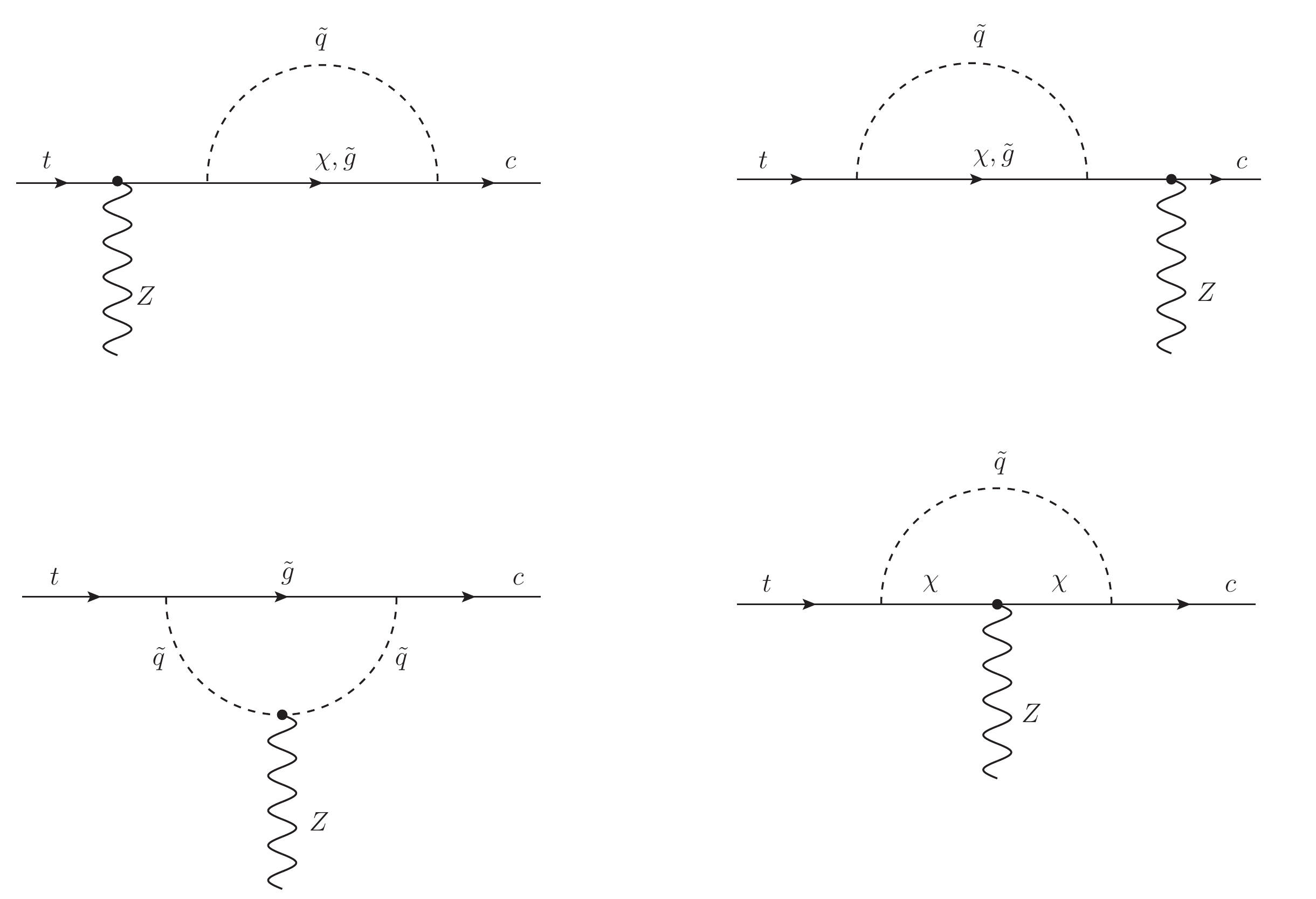}
\caption{\label{fig:Feynman-Diagrams-Top}Feynman diagrams for the
process $t\rightarrow cZ$. The dashed lines represent squarks exchange.}
\end{figure}

In the computations of the branching ratio
$\mathcal{B}\left(t\rightarrow cZ\right)$ we follow the
analysis of \cite{deDivitiis:1997sh} (see also~\cite{Li:1993mg}). We first work in the
basis where the squarks mass matrix is diagonal. In this basis
the diagrams relevant for the $t\rightarrow cZ$ process are
shown in Fig. (\ref{fig:Feynman-Diagrams-Top}).

The effective vertex relevant for FCNC can be parameterized as
\beq \label{l1}
-i\bar{u}(p)\left[P_{R} \left(F_{L}^{a}
q^{2}\gamma^{\mu}+F_{L}^{b}\slashed{q}q^{\mu}+
G_{L}i\sigma^{\mu\nu}q_{\nu}\right) +
P_{L}\left(F_{R}^{a}q^{2}\gamma^{\mu}+F_{R}^{b}\slashed{q}
q^{\mu}+G_{R}i\sigma^{\mu\nu}q_{\nu}\right)\right]\epsilon_{\mu}u(p+q)\,,
\eeq
where $P_{L,R}=\frac{\left(1\mp\gamma_{5}\right)}{2}$. The form
factors $F^{a}$ can be in general written as
\begin{eqnarray}
F_{L}^{a} & = & \frac{g_{s}^{2}}{4\pi^{2}}\sum_{\alpha,\beta=1}^{6}\left[K_{\tilde{c}_{L},\beta}^{\dagger}F_{1L}^{a}\left(\alpha,\beta\right)K_{\alpha,\tilde{t}_{L}}-K_{\tilde{c}_{L},\beta}^{\dagger}F_{2L}^{a}\left(\alpha,\beta\right)K_{\alpha,\tilde{t}_{R}}\right]\,,\\
F_{R}^{a} & = &
\frac{g_{s}^{2}}{4\pi^{2}}\sum_{\alpha,\beta=1}^{6}\left[K_{\tilde{c}_{R},\beta}^{\dagger}F_{1R}^{a}\left(\alpha,\beta\right)K_{\alpha,\tilde{t}_{R}}-K_{\tilde{c}_{R},\beta}^{\dagger}F_{2R}^{a}\left(\alpha,\beta\right)K_{\alpha,\tilde{t}_{L}}\right]\,,\end{eqnarray}
where the indices $\alpha,\beta$ identify the squarks mass
eigenstates and $K_{\alpha\beta}$ is the matrix that
diagonalizes the squarks mass matrix. Moreover, it can be shown
that analogous expressions are valid for $F^{b}$ and $G$ from
Eq.~\eqref{l1}.

Given that we perform a model independent analysis, we choose
the model parameters in order to obtain a robust bound on
$\left(\delta_{23}\right)_{LL}$. Thus we set
$\left(\delta_{23}\right)_{RL}=0$ in the squarks mass matrix.
In this case the only contribution to
$\left(\delta_{23}\right)_{LL}$ comes from
\begin{eqnarray}
F_{1L}^{a}\left(\alpha,\beta\right)  =a_{L}^{qqZ}
\delta_{\alpha\beta}\,C_{F}\,\frac{1}{2q^{2}}\left[B_{x}\left(m_{t}^{2},m_{\tilde{g}}^{2},m_{\alpha}^{2}\right)+B_{0}\left(m_{t}^{2},m_{\alpha}^{2},m_{\tilde{g}}^{2}\right)\right]\label{eq:Form_factor_relevant}\,,\end{eqnarray}
where $m_{t}$, $m_{\tilde{g}}$ and $m_{\alpha}$ correspond
respectively to the top, gluino and squarks masses. Furthermore
we introduce the definitions
$C_{F}=\frac{N^{2}-1}{2N}=\frac{4}{3}$ and
\begin{eqnarray}
a_{L}^{qqZ} =
\frac{g_{w}}{2\cos\left(\theta_{w}\right)}\left[-1+\frac{4}{3}\sin^{2}\left(\theta_{w}\right)\right] \,.
\end{eqnarray}
The complete expression for the form factors, including the
expressions for the decay into photons and gluons can be found
in \cite{deDivitiis:1997sh}. The quantities $B_{x}$ and $B_{0}$
in (\ref{eq:Form_factor_relevant}) are given by\begin{eqnarray}
B_{0}\left(p^{2},m_{1}^{2},m_{2}^{2}\right) & = & \mbox{pole terms}-\int_{0}^{1}d\alpha\:\log\left[p^{2}\alpha^{2}+\left(m_{1}^{2}-m_{2}^{2}-p^{2}\right)\alpha+m_{2}^{2}\right]\,,\nonumber\\
B_{x}\left(p^{2},m_{1}^{2},m_{2}^{2}\right) & = & \mbox{pole
terms}+\int_{0}^{1}d\alpha\:\alpha\log\left[p^{2}\alpha^{2}+\left(m_{1}^{2}-m_{2}^{2}-p^{2}\right)\alpha+m_{2}^{2}\right]\,,\end{eqnarray}
where the divergent parts in the above integrals cancel in the
final result. In the following we work in the approximation of
quasi-degeneracy for the squarks, thus moving to the super-CKM
basis. Expanding in terms of the mass insertions
$\left(\delta_{ij}^{u}\right)_{LL}$, we arrive at the following
expression for the part of the form factors contributing to
flavor violation
\beq
\left(F_{L}\right)_{ij}\big|_{q^2=m_Z^2}=\frac{g_{s}^{2}}{4\pi^{2}}a_{L}^{qqZ}C_{F}\frac{1}{2m_Z^{2}}f(x_t,x_g)\left(\delta_{ij}^{u}\right)_{LL}
\,,
\eeq
with $f(x_t,x_g)$ given by
\beq
\begin{split}
f(x_t,x_g)\equiv\frac{1}{2x_{t}^{2}h(x_t,x_g)}&\biggl\{h(x_t,x_g)\left[(x_{g}+x_{t}-1)\log(x_{g})-2x_{t}\right]\\
&+2\sqrt{h(x_t,x_g)}\left[\left(x_{g}-2\right)x_{g}+\left(x_{t}-1\right){}^{2}\right]\tan^{-1}\left(\frac{x_{g}-x_{t}-1}{\sqrt{h(x_t,x_g)}}\right)\\
&\left.-2\sqrt{h(x_t,x_g)}\left[\left(x_{g}-2\right)x_{g}+\left(x_{t}-1\right){}^{2}\right]\tan^{-1}\left(\frac{x_{g}+x_{t}-1}{\sqrt{h(x_t,x_g)}}\right)\right\}\,,
\end{split}
\eeq
and where we use
\beq
x_{g}\equiv\frac{m_{\tilde{g}}^{2}}{\tilde{m}_{Q}^{2}}\,,\quad
x_{t}\equiv\frac{m_{t}^{2}}{\tilde{m}_{Q}^{2}} \,, \quad
h(x_t,x_g)\equiv 2x_{g}(x_{t}+1)-x_{g}^{2}-(x_{t}-1)^{2} \,.
\eeq

In terms of the form factors, the expression for the branching
ratio, normalized to $\Gamma\left(t\rightarrow b+W\right)$, is
given by
\beq
\mathcal{B}\left(t\rightarrow
cZ\right)=\frac{1}{\Gamma\left(t\rightarrow
b+W\right)}\frac{1}{32\pi}\frac{m_{t}^{2}-m_{Z}^{2}}{2m_{t}^{3}}
\left.\left(F_{L}^{a}\right)^{2}\right|_{q^{2}=m_{Z}^{2}}\left(2m_{t}^{4}m_{Z}^{2}
+2m_{t}^{2}m_{Z}^{4}-4m_{Z}^{6}\right)\,.\label{eq:BR_tcZ}
\eeq
Evaluating Eq.~\eqref{eq:BR_tcZ} at
$\tilde{m}_{Q}=100\:\mbox{GeV}$ and $\tilde{m}_{Q}=
m_{\tilde{g}}$, we get the following bound for
$\left(\delta_{23}^{u}\right)_{LL}$
\beq
\left(\delta_{23}^{u}\right)_{LL} < 0.84
\,.\label{eq:Bound_delta_top}
\eeq

\subsection{Bottom Decay}

We now move to discuss the $b\rightarrow s\ell^{+}\ell^{-}$
transition. The branching ratio is given in terms of the Wilson
coefficients by~\cite{Fox:2007in}
\beq
\begin{split}
\mathcal{B}\left(B\rightarrow X_{s}\ell^{+}\ell^{-}\right)_{1<q^{2}<6\;\mathrm{GeV}^{2}}=
 &10^{-6}\left\{ 1.55+35100\left[\left|\Delta C_{9}\left(m_{W}\right)\right|^{2}+\left|\Delta C_{10}\left(m_{W}\right)\right|^{2}\right.\right.\label{eq:BR_Xll}\\
 &\left.\left.{}+\mbox{Re}\left[\left(180+5i\right)\Delta C_{9}\left(m_{W}\right)\right]\right]-360\mbox{Re}\left[\Delta C_{10}\left(m_{W}\right)\right]\right\}\,.
\end{split}
\eeq

The contributions to flavor violation coming from the MSSM can
be derived in the MI approximation. In order obtain a robust
bound on $\left(\delta_{23}^{d}\right)_{LL}$, we neglect the
chargino contributions, which depend on additional parameters,
such as $\mu$, $\tan \beta$ etc.. Under this assumption, the
explicit values for the MSSM contributions to $C_{9}$ with
$\tilde{m}_{Q}=100\:\mbox{GeV}$ and $\tilde{m}_{Q}=
m_{\tilde{g}}$ are
\beq
\Delta C_{9}\left(m_{W}\right)=
-1.75\left(\delta_{23}^{d}\right)_{LL}\label{eq:DelC9_DelC10}
\,,
\eeq
and $C_{10}$ vanishes, see~\cite{Lunghi:1999uk}\footnote{Note
that there is a factor of 2 between the definitions of the
operator $O_9$ in~\cite{Lunghi:1999uk} and~\cite{Fox:2007in}}
(and similar Refs.~\cite{Buchalla:2000sk}).

Combining Eqs.~\eqref{eq:BR_Xll} and~\eqref{eq:DelC9_DelC10}
with the experimental bound in Eq.~\eqref{tbdecay}, we obtain
the following bound
\beq
\left(\delta_{23}^{d}\right)_{LL}<0.003
\,.\label{eq:Bound_delta_bottom}
\eeq

\subsection{Best alignment}
As a result of the large difference between the top and bottom
bounds, Eqs.~\eqref{eq:Bound_delta_top}
and~\eqref{eq:Bound_delta_bottom}, the best alignment scenario
is practically equivalent to alignment with the down sector. In
this case, $\left(\delta_{23}^d\right)_{LL}=0$ and
$\left(\delta_{23}^u\right)_{LL}$ is simply proportional to the
squarks mass squared difference multiplied by $V_{cb}
V^*_{tb}$. Taking for concreteness $\tilde m_Q=\left(2
m_{\tilde Q_2}+m_{\tilde Q_3}\right)/3$ (appropriate for models
with only weak degeneracy~\cite{Raz:2002zx}), the bound is then
\beq
\frac{\left|m_{\tilde Q_2}^2- m_{\tilde Q_3}^2\right|}{\left(2
m_{\tilde Q_2}+m_{\tilde Q_3}\right)^2}
 <20\left(\frac{\tilde m_{Q}}{100\mbox{\,GeV}}\right)^2\,.
\eeq
We find therefore that no significant constraint on the level
of degeneracy can be obtained.

\subsection{$\Delta F=2$ Processes}
We next move to describe $\Delta F=2$ processes, for which the
results of Sec.~\ref{sec:uutt} are easily applied. Considering
again the leading order in the expansion
$\left(\delta_{13}\right)_{LL}$, we arrive at the following
expression for the length of $X_Q$
\beq
L=\frac{\alpha_{s}}{18}\,\sqrt{g(x)\over2} \,
\left(\delta_{13}\right)_{LL}\,,
 \eeq
where $x=m_{\tilde{g}}^{2}/\tilde m_{Q}^{2}$ and $g(x)$ is a
known kinematic function~\cite{Ciuchini:1998ix}. Taking $\tilde
m_Q = 100\mbox{\,GeV}$ and $m_{\tilde{g}}= \tilde m_{{Q}}$,
which implies $g(1)=1$, we find
\beq
\frac{\left|m_{\tilde Q_1}^2- m_{\tilde Q_3}^2\right|}{\left(2
m_{\tilde Q_1}+m_{\tilde Q_3}\right)^2}
 <0.45\left(\frac{\tilde m_{Q}}{100\mbox{\,GeV}}\right)^2\,.
\eeq
It should be mentioned that, by carefully tuning the squark and
gluino masses, one finds a ``sweet spot'' in parameter space,
where an even weaker bound is obtained~\cite{Crivellin:2010ys}.

\section{Warped Extra Dimension} \label{sec:rs}
The next framework that we analyze is the Randall-Sundrum (RS)
warped extra dimension~\cite{Randall:1999ee}. Here we consider
a generic form of this theory, in which all matter fields
propagate in the bulk, while the Higgs field is confined to the
IR brane. The localization of the fermions in the bulk
generates mass hierarchies and mixing angles, thus addressing
the flavor
puzzle~\cite{Davoudiasl:1999tf,Pomarol:1999ad,Grossman:1999ra,
Gherghetta:2000qt}. Moreover, there is an inherent mechanism in
this framework which provides protection against large FCNCs,
namely RS-GIM~\cite{Agashe:2004cp}.

The $\Delta F=2$ operator in Eq.~\eqref{o1} is most dominantly
induced by a tree level Kaluza-Klein (KK) gluon exchange.
Focusing on this contribution, we can write
\beq \label{rs2_expression}
m_{\rm KK}=\Lambda_{\rm NP} \, , \quad \xqt\cong
\frac{g_{s*}}{\sqrt{6}} \,\mathrm{diag}
(f^2_{Q^1},f^2_{Q^2},f^2_{Q^3}) \, ,
\eeq
before removing the trace, where $g_{s*}$ is the dimensionless
5D coupling of the gluon ($g_{s*} \approx 3$ after one loop
matching~\cite{Agashe:2008uz}) and the $f_{Q^i}$'s are the
values of the quark doublets on the IR brane. These are related
to each other through the CKM elements~-- $f_{Q^1,Q^2}/f_{Q^3}
\sim V_{ub},V_{cb}$. Plugging the length of $\xqt$ calculated
from Eq.~\eqref{rs2_expression} into Eq.~\eqref{dd_bound}, the
resulting limit is
\beq \label{rs2_bound}
m_{\rm KK}>0.4 f^2_{Q^3} \; \mathrm{TeV} \, ,
\eeq
where $f_{Q^3}$ is typically in the range of 0.4-$\sqrt{2}$.

The $\Delta F=1$ operator in Eq.~\eqref{3g_operator} is
generated, among others, via mixing between the zero mode (SM)
Z boson and its KK excitations. The bulk profile of these
higher modes is localized near the IR brane, which results in
non-universal couplings to the fermions. This in turn generates
flavor violating couplings in the mass
basis~\cite{Agashe:2004cp, Agashe:2006wa}, roughly given by
\beq
\delta g_Z \cong \log
\left(\frac{M_{\mathrm{Pl}}}{\mathrm{TeV}} \right) \left(
\frac{m_Z}{m_{\rm KK}} \right)^2 \, ,
\eeq
where $M_{\mathrm{Pl}}$ is the Planck scale. We focus only on
this contribution, as the others are of the same
order~\cite{Agashe:2006wa} (see also~\cite{Bauer:2009cf} for a
recent discussion on RS flavor violation in the up sector), and
write
\beq
\xqo \cong g_{Z*}\, \delta g_Z\,
\mathrm{diag}(f^2_{Q^1},f^2_{Q^2},f^2_{Q^3}) \, ,
\eeq
with $g_{Z*}$ as the dimensionless 5D coupling of the Z to
left-handed up type quarks ($g_{Z*} \cong 1.2$ at one loop).
The bound that stems from this via case (ii) of
Eq.~\eqref{3g_bounds} is
\beq \label{rs1_bound}
m_{\rm KK}>0.33 f^2_{Q^3} \; \mathrm{TeV} \, .
\eeq

The constraints presented in Eqs.~\eqref{rs2_bound}
and~\eqref{rs1_bound} are rather weak, compared to known limits
on RS, but they are immune to various models of
alignment~\cite{RSalignment}.

\section{Conclusions \& Outlook} \label{sec:conc}
The field of flavor physics has arrived to a point where it is
clear that flavor and CP violation are dominated by the
standard model Cabibbo-Kobayashi-Maskawa mechanism. However, at
this time we cannot determine what will be the nature of the
expected, yet to be discovered, new dynamics at the TeV scale.
Our current indirect experimental data is certainly not mature
enough to point to a flavor blind dynamics. However, if the new
physics is non-universal, we can, with reasonable certainty,
expect that, if accessible to the LHC, it would share the
approximate symmetry structure of the SM. Thus flavor should be
dominantly broken via the third generation sector~\cite{NMFV}.
It looks, therefore, useful to derive a flexible TeV effective
description for flavor violation, that allows to manifestly
incorporate the standard model form of flavor breaking. Our
covariant formalism enables us not only to describe generic new
physics in a model independent manner, but also to naturally
describe the SM breaking pattern, given that the formalism's
basic building blocks are the SM sources of flavor breaking.

We find that projected LHC bounds on $\Delta t=1$ processes
lead to a new model independent constraint on the strength of
left-handed quarks flavor violation, even in the presence of
general flavor alignment mechanisms. The projected bound on
$\Delta t=2$ transitions from same sign tops production at the
LHC is also studied. In this case we find a stronger bound due
to recent experimental constraint on CP violation in
$D-\overline D$ mixing. We use our analysis to obtain new
limits on supersymmetric and warped extra dimension models of
alignment, which turn out to be rather weak, but nonetheless
replacing practically non-existing current bounds.

In our study we have only focused on the leading framework
independent contributions. We want to point out that in our
$\Delta F=1$ analysis we have only considered the contribution
from the operator $O_{LL}^h$, given that $O_{LL}^u$ induces
$b\to s$ transitions only at one loop. Nevertheless, it is
interesting to analyze the bounds induced by the latter
operator. Moreover, the supersymmetric contributions that we
have considered above are actually not coming from
$O_{LL}^h$~\cite{Buchalla:2000sk}, and in any case were found
to be very weak. It is therefore worthwhile to study the role
of other contributions mediated by the Higgs and chargino
sector, depending on additional parameters. It is also worth
mentioning that the set of operators discussed above lead to
other flavor violating processes, such as $b\to s\nu \bar \nu$,
$b\to \mu \bar \mu$ etc., which were not analyzed by us.
However, given that the constraints coming from top flavor
violation were always much weaker, inclusion of other processes
would not change the qualitative nature of our results, yet
interesting to further investigate. In addition, one could
derive a projected bound on our $\Delta t=1$ operators by
studying the cases where on-shell top and $Z$ are obtained in
the final state (for a recent work along these lines, but which
considered different set of operators see
\eg~\cite{AguilarSaavedra:2010rx} and references therein). We
expect that the resulting bounds would be
weaker~\cite{Fox:2007in}, yet to the best of knowledge a
dedicated study of this matter does not exist at present.

Another interesting framework that can easily be explored using
our formalism is the minimal flavor violation (MFV)
scenario~\cite{MFV}. In this case the new flavor breaking
source is just a function of the SM Yukawa matrices. It is
possible to generate large top flavor violation within this
framework~\cite{Kagan:2009bn} and also $b\to s$
transitions~\cite{MFV}. We expect that the weakest possible
bound in this case would be similar to what we derived in the
general scenario. The new physics source would just be an
appropriate linear combination of $Y_u Y_u^\dagger$  and $Y_d
Y_d^\dagger$. This, however, only corresponds to a narrow
subclass of MFV models, denoted as linear
MFV~\cite{Kagan:2009bn}. In covariant language, the linear MFV
limit simply corresponds to cases where the flavor violating
sources reside on the $Y_u Y_u^\dagger-Y_d Y_d^\dagger$ plane.
In general, as our covariant basis explicitly demonstrates, an
arbitrary function of the Yukawa matrices could produce any
kind of flavor and CP
violation~\cite{Colangelo:2008qp,Mercolli:2009ns,
Ellis:2009di}. Nonetheless, higher powers of the Yukawas
approach the $U(2)$ limit. Consequently, the general MFV case
would be approximately characterized by new physics sources
belonging to the submanifold generated in this limit, described
in detail in the text. The non-linear limit of MFV is typically
obtained in models with large anomalous dimensions or large
logs, where third generation Yukawa resummation is required to
obtain the low energy effective theory~\cite{Kagan:2009bn,
Isidori:2010kg}. Thus, distinguishing between these two
limiting cases of MFV models could yield precious information
on microscopic type of new physics, well beyond the reach of
the LHC.

\vspace*{1cm}

\section*{Acknowledgements}
GP is the Shlomo and Michla Tomarin career development chair.
The work of GP is supported by the Israel Science Foundation
(grant \#1087/09) EU-FP7 Marie Curie, IRG fellowship and the
Peter \& Patricia Gruber Award.

\appendix
\section{The Number of Different Adjoints in a $\Delta F=2$ Operator}
\label{app:adj}
The analysis above for $\Delta F=2$ operators is based on the
assumption that these are all written as in Eq.~\eqref{o1}.
However, one might wonder whether it is possible to consider
this form with two different adjoints, that is $O_1 \sim \left(\,
\overline{Q}_{i} (X_Q)_{ij} \gamma_\mu Q_{j} \right) \left(
\overline{Q}_{i} (Y_Q)_{ij} \gamma^\mu Q_{j} \right)$ with $X_Q
\neq Y_Q$. This means that the new physics is described by two
objects, which complicates the analysis significantly.

To address this issue, we appeal to group theory. For three
generations, the two quarks in the operator form together a
$\mathbf{6}$ representation of the SU(3)$_Q$ flavor group, and
similarly the two anti-quarks are in a $\overline{\mathbf{6}}$.
The entire operator together is then part of the reducible
representation $\mathbf{6} \otimes \overline{\mathbf{6}}$,
which decomposes to $\mathbf{1} \oplus \mathbf{8} \oplus
\mathbf{27}$. Clearly, the operator is therefore described by
the symmetric $\mathbf{27}$ irreducible representation.

Since we are interested in quark structures such as $\bar{t}c
\bar{t}c$ etc., the state inside the $\mathbf{27}$
representation is of maximal weight. Accordingly, this operator
is uniquely defined with a single adjoint object. In other
words, if we use two different adjoints, $X_Q$ and $Y_Q$, as
suggested above, then only specific linear combination of their
components would appear in physical observables, such that they
can be absorbed into a single adjoint. This argument of course
applies also to the two generations case, for which the
operator resides in the $\mathbf{5}$ (spin 2) representation of
SU(2).

\section{Complete Covariant Basis - Explicit Decomposition} \label{app:decomp}
In sec.~\ref{sec:3g_full} we construct a complete basis for
three generations flavor space. In order to give a sense of the
physical interpretation of the different basis elements, we
present here their decomposition in terms of Gell-Mann
matrices, in the down mass basis (writing only the dependence
of the leading terms on $\lmc$ and omitting for simplicity
$\mathcal{O}(1)$ factors such as the Wolfenstein parameter
$A$). First we show the simpler case of taking the CKM phase to
zero, which yields
\beq
\begin{split}
\hat D_1 \sim& \left\{-1,0,0,0,0,0,0,0\right\} \, , \\ \hat D_2
\sim& \left\{0,-1,0,0,0,0,0,0\right\} \, ,\\ \cdn \sim&
\left\{2\lmc,0,1,0,0,0,0,0\right\} \, ,\\
\hat D_4 \sim& \left\{0,0,0,-1,0,-\lmc,0,0\right\} \, ,\\ \hat
D_5 \sim& \left\{0,0,0,0,-1,0,-\lmc,0\right\} \, ,\\ \hjd^n
\sim& \left\{0,0,0,-\lmc,0,1,0,0\right\} \, ,\\ \hj^n \sim&
\left\{0,0,0,0,-\lmc,0,1,0\right\} \,, \\ \hadn =&
\left\{0,0,0,0,0,0,0,-1\right\} \, ,
\end{split}
\eeq
where the values in each set of curly brackets stand for the
$\Lambda_1, \ldots ,\Lambda_8$ components. This shows which
part of an object each basis element extracts under a dot
product, relative to the down sector. For instance, $\hat D_1$
is proportional to $\Lambda_1$, and therefore represents the
real part of a $2 \to 1$ transition. Restoring the CKM phase,
we find
\beq
\begin{split}
\hat D_1 \sim& \left\{-1,\eta,0,0,0,0,0,0\right\} \, , \\ \hat
D_2 \sim& \left\{-\eta,-1,0,0,0,0,0,0\right\} \, ,\\ \cdn \sim&
\left\{2\lmc,-2\eta\lmc,1,0,0,0,0,0\right\} \, ,\\
\hat D_4 \sim& \left\{0,0,0,-1,\eta,-\lmc,-\eta\lmc^3,0\right\} \, ,\\
\hat D_5 \sim& \left\{0,0,0,-\eta,-1,\eta\lmc^3,-\lmc,0\right\} \, ,\\
\hjd^n \sim& \left\{0,0,0,-\lmc,\eta\lmc,1,\eta\lmc^2,0\right\} \, ,\\
\hj^n \sim& \left\{0,0,0,-\eta\lmc,-\lmc,-\eta\lmc^2,1,0\right\} \,, \\
\hadn =& \left\{0,0,0,0,0,0,0,-1\right\} \, ,
\end{split}
\eeq
where $\eta$ is the CPV Wolfenstein parameter. Finally, the
leading term decomposition of $\au$ in the down mass basis is
\beq \label{au_gellman_decomp}
\au \sim \left\{ -\lmc y_c^2- \lmc^5 yt^2, \eta\lmc^5 y_t^2,
-(y_c^2+\lmc^4 y_t^2)/2, \lmc^3 y_t^2,-\eta\lmc^3 y_t^2,-\lmc^2
y_t^2, -\eta \lmc^4 y_t^2,-y_t^2/\sqrt3 \right\} \, ,
\eeq
neglecting the mass of the up quark. It is interesting to
notice the differences from Eq.~\eqref{au_cov_decomp}, where
$\au$ is decomposed in the covariant basis. For instance, it is
well known that CPV in $s \to d$ transitions within the SM is
produced only through mixing with the third generation, hence
$\au \cdot \Lambda_2$ is suppressed by $\eta\lmc^5 y_t^2$ in
Eq.~\eqref{au_gellman_decomp}. This is in contrast to $\au
\cdot \hat D_2 \sim \lmc y_c^2$ in Eq.~\eqref{au_cov_decomp}.

\section{Bounds from the First Two Generations} \label{app}
The three generations framework presented in Sec.~\ref{sec:3g}
is oriented at evaluating flavor violating interactions of the
bottom and the top. This is manifest from the approximation of
neglecting the masses of the first two generation quarks.
However, any new physics contribution need not respect this
symmetry in general, thus it may lead to flavor violation
between the first two generations.

An interesting example for this is the case of $\xqo=L\, \hjq$,
mentioned in Sec.~\ref{sec:app}, which represents full
alignment for both sectors regarding third generation decays.
Yet it does induce flavor violation between the first two
generations, so we can use experimental limits to constrain
this form of $\xqo$.

First note that $\hjq$ does not commute with $\ad$ and $\au$
anymore, once all quark masses are restored. Thus $\xqo$ can
maintain this form (written explicitly as $L \,(\sqrt3
\Lambda_3+\Lambda_8)/2$) only in a specific basis. For
simplicity, we take it to be in the down mass basis, such that
it only induces flavor violation in the up sector, where the
constraints are weaker (we avoid tweaking the alignment between
the sectors to obtain the weakest bound, since it would only
slightly change the result below). Therefore, we need to
consider the contribution of $\xqo$ to $c \to u$ decays. This
can be calculated using the complete covariant basis defined in
Sec.~\ref{sec:3g}. To do this, we need to construct that basis
around the up sector (by replacing $\ad \leftrightarrow \au$ in
the entire derivation). The coefficient of the operator in
Eq.~\eqref{3g_operator} is thus given by
\beq \label{jq_cu}
\xqo=L \hjq \ \longrightarrow \ \left(C_{LL}^h\right)_{c \to
u}=\left| \left( \xqo \cdot \hat D_1 \right)^2+ \left( \xqo
\cdot \hat D_2 \right)^2 \right|=0.38\, L \, .
\eeq

For the experimental constraint, we take the result given in
Eq.~(28) of Ref.~\cite{Fajfer:2007dy}, written in terms of the
standard Wilson coefficients. Using the relation between
$C_{LL}^h$ and the Wilson coefficients~\cite{Fox:2007in}, we
find
\beq \label{cu_exp_bound}
\left(C_{LL}^h\right)_{c \to u}<1.05 \ltev^2 \, .
\eeq
Plugging Eq.~\eqref{jq_cu} into~\eqref{cu_exp_bound} results in
\beq
L<0.59 \ltev^2; \quad  \Lambda_{\rm NP}>1.7 \, \mathrm{TeV} \,
.
\eeq

The above example represents a leading order contribution of
new physics to the $2 \to 1$ transition. At the end of
Sec.~\ref{sec:app} we consider a form of $\xqo$ designed to
avoid such contributions, yet it still generates this process
at higher order in $\lambda_{\rm C}$. For instance, the
effective CKM matrix as a 2-3 rotation takes $\xqo=L \, \had$
(complete alignment with the down sector) into a combination of
$\had$, $\hat C_d$ and $\hjd$, none of which involves
transitions among the first two generations. However, if we use
the full CKM, as required when all the masses are accounted
for, then we do get a contribution to $c \to u$ at
$\mathcal{O}(\lambda_{\rm C}^5)$, given by
\beq
\xqo=L \had \ \longrightarrow \ \left(C_{LL}^h\right)_{c \to
u}=\left| \left( \xqo \cdot \hat D_1 \right)^2+ \left( \xqo
\cdot \hat D_2 \right)^2 \right|=2.8 \times 10^{-4} L \,.
\eeq
Plugging this into Eq.~\eqref{cu_exp_bound}, the resulting
bound is $L<800$ for $\Lambda_{\rm NP}=1$ TeV. This is more
than two orders of magnitude weaker than the limit from top
decay (case (i) in Eq.~\eqref{3g_bounds}), and the
corresponding bound for the optimal alignment (case (ii)) is of
the same order. Hence this suppression is enough to make this
bound irrelevant, as compared to the one based on the future
LHC top measurements.

Finally, it is instructive to see explicitly the contribution
of the covariant basis elements to flavor violation in the
first two generations. In the down mass basis, say, these
elements can be identified as
\beq
\begin{split}
\had&=-\Lambda_8 \, , \quad \hj=\Lambda_7\,, \quad
\hjd=\Lambda_6\,, \quad
\hat C_d=\Lambda_3 \,, \quad \hD=\Lambda_{1,2,4,5} \,, \\
\hadp&=\frac{\Lambda_3-\sqrt3
\Lambda_8}{2}=\mathrm{diag}(0,-1,1)\,, \quad \hjq=\frac{\sqrt3
\Lambda_3+\Lambda_8}{2}= \frac{1}{\sqrt3}\mathrm{diag}(2,-1,-1)
\, .
\end{split}
\eeq
Clearly, only $\Lambda_{1,2}$ generate the transition $s \to
d$. When a full CKM rotation is applied to move to the up mass
basis, the following contributions to $c \to u$ arise:
\begin{itemize}
\item The $\Lambda_{1,2}$ components of $\hD$ yield a direct
    contribution.
\item $\hat C_d$, $\hadp$ and $\hjq$ all contain
    $\Lambda_3$, thus producing an
    $\mathcal{O}(\lambda_{\rm C})$ contribution.
\item The $\Lambda_{4,5}$ components of $\hD$ generate this
    contribution via a 2-3 rotation, which is at
    $\mathcal{O}(\lambda_{\rm C}^2)$.
\item $\hj$ and $\hjd$ require a 1-3 rotation, of
    $\mathcal{O}(\lambda_{\rm C}^3)$.
\item For $\had$, a combination of a 2-3 and a 1-3 CKM
    rotation is needed, which is at
    $\mathcal{O}(\lambda_{\rm C}^5)$.
\end{itemize}

In Sec.~\ref{sec:app} we throw away the $\hD$ component of
$\xqo$ (among others), claiming that it contributes to $2 \to
1$ at leading order. Here we see that actually the
$\Lambda_{4,5}$ parts of $\hD$ are only relevant at
$\mathcal{O}(\lambda_{\rm C}^2)$. Nonetheless, these do
contribute to both the top and the bottom decays. Then, since
under the CKM transformation they mostly mix with
$\Lambda_{1,2}$, minimizing their contribution requires large
$\Lambda_{1,2}$ components, which in turn directly produce $2
\to 1$.

\end{document}